\documentstyle[12pt,aasms4]{article}

\received{}
\accepted{}
\journalid{}{}
\articleid{}{}

\slugcomment{Accepted by {\it The Astrophysical Journal}}

\lefthead{Mirabal {\it et al.}}
\righthead{Search for 3EG J1835+5918}

\begin{document}

\def\etal{{\it et al.}}
\def\eg{{\it e.g.}}
\def\ie{{\it i.e.}}
\def\vs{{\it vs.}}
\def\etc{{\it etc.}}
\def\kms{km~s$^{-1}$}
\def\Msol{M$_\odot$}
\def\lsim{\mathrel{\lower .85ex\hbox{\rlap{$\sim$}\raise
.95ex\hbox{$<$} }}}
\def\gsim{\mathrel{\lower .80ex\hbox{\rlap{$\sim$}\raise
.90ex\hbox{$>$} }}}

%
%
\def\pz{\phantom{0}}
\def\lsim{\mathrel{\lower .85ex\hbox{\rlap{$\sim$}\raise
.95ex\hbox{$<$} }}}
\def\gsim{\mathrel{\lower .80ex\hbox{\rlap{$\sim$}\raise
.90ex\hbox{$>$} }}}
\def\bv{($B-V$)}
\def\vr{($V-R$)}
\def\br{($B-R$)}
\def\ub{($U-B$)}
\def\vi{($V-I$)}
\def\ri{($R-I$)}
\def\source{3EG~J1835+5918}
\def\ro{{\it ROSAT\/}}
\def\asca{{\it ASCA\/}}

\title{Search for the Identification of 3EG J1835+5918: Evidence for
a New Type of High-Energy Gamma-ray  Source}


\author{N. Mirabal and J. P. Halpern}
\affil{Astronomy Department, Columbia University, 550 West 120th Street,
 New York, NY 10027}
\affil{abulafia@astro.columbia.edu,jules@astro.columbia.edu}
\authoremail{abulafia@astro.columbia.edu, jules@astro.columbia.edu}
\author{M. Eracleous}
\affil{Department of Astronomy and Astrophysics, The Pennsylvania State
University, \\525 Davey Laboratory, University Park, PA 16802}
\affil{mce@astro.psu.edu}
\author{R. H. Becker}
\affil{Physics Department, University of California, Davis, CA  95616}
\affil{bob@igpp.llnl.gov}

\begin{abstract}
\rightskip 0pt \pretolerance=100 \noindent
The EGRET source \source\ is the brightest and most accurately
positioned of the as-yet unidentified high-energy $\gamma$-ray sources
at high Galactic latitude ($\ell,b=89^{\circ},25^{\circ}$).
We present a multiwavelength study of the region around it,
including X-ray, radio, and optical imaging surveys, as well as optical
spectroscopic classification of most of the active objects in this area.
Identifications are made of all but one of the {\it ROSAT}
and {\it ASCA} sources in this region to a flux limit of approximately
$5 \times 10^{-14}$~erg~cm$^{-2}$~s$^{-1}$, which is $10^{-4}$
of the $\gamma$-ray flux.  The identified
X-ray sources in or near the EGRET error ellipse are radio-quiet
QSOs, a galaxy cluster, and coronal emitting stars.
We also find eight quasars
using purely optical color selection, and we have
monitored the entire field for variable optical objects on short and
long time scales without any notable discoveries.
The radio sources inside the error ellipse are all fainter than 
4~mJy at 1.4~GHz.
There are no flat-spectrum radio sources in the vicinity; the
brightest neighboring radio sources are steep-spectrum radio galaxies 
or quasars.  Since no blazar-like or pulsar-like candidate has
been found as a result of these searches,
\source\ must be lacking one or more of the physically
essential attributes of these known classes of $\gamma$-ray emitters.
If it is an AGN it lacks the beamed radio emission of blazars by at least
a factor of 100 relative to identified EGRET blazars. If it is an isolated
neutron star, it lacks the steady thermal X-rays from a cooling surface
and the magnetospheric non-thermal X-ray emission that is characteristic
of all EGRET pulsars.  If a pulsar, \source\
must be either older or more distant than Geminga,
and probably an even more efficient or beamed $\gamma$-ray engine.
One intermittent \ro\ source falls on a blank optical
field to a limit of $B > 23.4, V > 23.3$, and 
$R > 22.5$. In view of this conspicuous absence, RX~J1836.2+5925 should
be examined further as a candidate for identification with \source\ and 
possibly the prototype of a new class of high-energy $\gamma$-ray source.

\end{abstract}

\keywords{gamma rays: observations --- pulsars: general -- 
radio continuum: galaxies --- X-rays: general}

\section{Introduction}

One of the most important advances in high-energy astrophysics in recent 
years is the discovery of 271 persistent 
high energy $\gamma$-ray sources by the EGRET instrument aboard
the Compton Gamma-ray Observatory ({\it CGRO\/},
Hartman \etal~1999).  While the detection of these sources is a major
success, identification of their nature and origin  
has turned out to be a more challenging task.  The principal method of
identification, which relies on statistical evidence that blazars are the
dominant population, is to find positional coincidences between EGRET sources 
and flat-spectrum radio/millimeter sources
(Thompson \etal~1995, 1996; Mattox \etal~1997; Bloom et al. 1997).
By definition blazars are 
flat-spectrum, radio-loud AGNs with polarized and variable optical emission. Although numerous efforts have been made at various wavelengths, only
about one third of all EGRET sources have been identified
with any degree of
confidence.  On the latest count these identifications include 66 
blazars, \ie, flat-spectrum radio quasars or BL~Lac objects
(Hartman \etal~1999), seven rotation-powered pulsars (Hartman 
\etal~1999, Kaspi \etal~2000, Ramanamurthy \etal~1995),
the nearby radio galaxy Cen A, and the Large Magellanic Cloud.
Therefore approximately 196 EGRET sources remain unidentified 
with roughly half of these located at high Galactic latitude, 
$b > 10^\circ$. 

Many difficulties attend the identification of EGRET sources close to the
Galactic plane, but even at high Galactic latitude, the size of the
typical error circle and the lack of a tight relation between 
gamma-ray flux and other properties such as X-ray flux and core radio
flux prevent all but the brightest counterparts from being identified 
securely on the basis of position alone. The  
absence of obvious counterparts also admits
the possibility that there is another
population with characteristics unlike the identified EGRET 
sources.  We have decided to explore the latter possibility by means of
detailed work at other wavelengths, while in the long term
the situation should improve
considerably with the next generation high-energy $\gamma$-ray mission
{\it GLAST\/}, which will produce more precise source locations.

We have chosen for a case study
the unidentified EGRET source \source. This object may be the best
candidate for the prototype of
a new population different from blazars or pulsars. 
It is the brightest of the as-yet unidentified EGRET sources
at high Galactic latitude ($\ell,b=89^{\circ},25^{\circ}$),
and the one with the
smallest error circle.  Because it is strongly detected
and well away from the confusing diffuse emission in the Galactic plane,
\source\ is localized to within a radius of only $12^{\prime}$
at 99\% confidence, which makes a deep multiwavelength search for
a counterpart feasible.  The latest analysis of the EGRET
observations of \source\ leads to the conclusion that it shows no strong
evidence for variability (Reimer \etal~2000).  Its spectrum can
be fitted by a power law of photon index --1.7 from 70~MeV to
4~GeV, with a turndown above 4~GeV.  Such temporal and spectral
behavior is more consistent with a rotation-powered pulsar
than a blazar.  Unlike \source, blazars
are highly variable, and exhibit steeper spectra.

Prior to the observations reported herein,
there were no known active galactic nuclei (AGNs)
or pulsars in the error circle of \source.
Examination of existing catalogs finds no flat-spectrum 
radio source (Mattox \etal~1997), no 1.4~GHz radio source of any type 
brighter than 4~mJy in the NRAO-VLA Sky Survey 
catalog (NVSS, Condon \etal~1998),
and no 4.85~GHz source brighter than 20~mJy (Becker, White, \& Edwards~1991).
Observations by Nice \& Sayer (1997)
find and no radio pulsar to an upper limit of 1~mJy at  
770~MHz.  Furthermore, all of the known gamma-ray blazars and
pulsars appear brighter in X-rays than the upper limit that we shall 
present for 
\source.  In light of these facts, \source\ cannot be a blazar 
unless it is a radio-quiet one (requiring a redefinition of this concept),
nor a pulsar unless, as we shall show, it 
is one with unprecedented characteristics.

In this paper we present the results of radio, X-ray, and 
optical observations of the location of \source. 
The outline of the paper is as 
follows: \S 2 describes our multiwavelength data acquisition and selection 
techniques. \S 3 describes the optical spectroscopy of candidates
and the overall results.
\S 4 details notable properties of individual objects and assesses
their prospects as the identification of \source.
Multiwavelength comparisons with known $\gamma$-ray sources 
are addressed in \S 5, and the implications and conclusions of our work
are discussed in \S\S 6 and 7.  

\section{Observations}

\subsection{Optical Photometry and QSO Candidate Selection}

The principal body of optical data for this study is a 
series of standard $UBV$ and Cousins $R$ CCD images of the error circle
of \source\ which we obtained using the MDM Observatory 1.3m telescope 
during a photometric run
in 1998 June and July.  A thinned, back-illuminated $2048 \times 2048$
SITe CCD was used to cover a $17^{\prime} \times  17^{\prime}$ field
with multiple exposures.
A mosaic of four such overlapping fields enabled us to observe a
$32^{\prime} \times  32^{\prime}$ region centered on the most likely
EGRET source position (B. Dingus, private communication).  Our images
thus cover the entire 99\% confidence region specified in the Third EGRET 
Catalog (Hartman \etal~1999), which can be approximated as an ellipse
of major axis $24^{\prime}$.  
In 1997 July we had covered the same field in the $V$
and $I$ bands only, and all of the $V$-band images were used to search for
variability on long (year) and short (hours to days) time scales.
The images were processed using standard IRAF/DAOPHOT procedures.
Approximately 5000 objects were measured inside a $15^{\prime}$ 
radius circle.
The photometry described here was calibrated using Landolt standard stars
(Landolt 1992).
Typical limiting detections achieved were $U=22.1, B=23.4, V=22.5$,
and $R=22.5$.
Galactic extinction in this field is small but not negligible; Schlegel, Finkbeiner, \& Davis (1998) give $E(B-V) = 0.045$, corresponding to 
$A_U = 0.25$, $A_B = 0.20$, $A_V = 0.15$, and $A_R = 0.12$.
Magnitudes quoted in this paper are observed, \ie, not corrected for extinction.

We derived a list of QSO candidates from this photometry using the
standard ultraviolet excess selection
technique.  Following Hall \etal~(1996),
we required plausible quasar candidates to have either $(B-V) < 0.4$ and
$(U-B) > -0.3$, or $(B-V) < 0.6$ and $(U-B) < -0.3$.
This selection is effective
in separating QSOs from the stellar locus, and is efficient in
detecting them out to $z = 2.2$ (Hall \etal~1996; Fan 1999).
We note that of the current identifications in the 3EG catalog,
which are unbiased by optical selection, the largest redshift is
only $z = 2.286$, and all of their optical counterparts are brighter than
$V = 22.1$.  Our color selection should also permit the discovery
of any object that has a power-law continuum, which produces a UV excess,
and especially a synchrotron spectrum which peaks above the optical band,
\eg, those blazars commonly referred to as high-energy peaked.
Thus, our technique is sensitive to most of the known EGRET
blazars, and useful to search for a UV excess counterpart that might
be expected on the basis of the absence of strong radio emission.

The major complication in this search comes in 
separating quasars from white dwarfs, blue field stars, and compact 
emission-line galaxies
which often have similar blue colors and
are known major contaminants of quasar color surveys.  Further criteria
can be applied using additional colors, but  
we decided to allow maximum freedom in the selection criteria in order
to avoid excluding possibly interesting candidates.  A total of 40 
such candidates to a limiting magnitude of $B = 21$ were selected for
follow-up spectroscopy.  In subsequent sections of this paper
we discuss the eight QSOs
that were discovered in our spectroscopic observations.

\subsection{X-ray Observations}

A total of three X-ray observations were made that cover the entire
99\% error ellipse of \source, two by the \ro\ High Resolution Imager (HRI)
and one by \asca.  The first \ro\ observation took place on 1995
February 2--4, with a total exposure time of 9,186~s.  Five 
point-like X-ray sources were detected in this image, which reached
a minimum detectable intrinsic
flux of $7.4 \times 10^{-14}$~erg~cm$^{-2}$~s$^{-1}$
in the 0.1--2.4~keV band, assuming a power-law spectrum with photon
index 2.0 and Galactic $N_{\rm H} = 4.6 \times 10^{20}$~cm$^{-2}$.
A longer HRI observation of the same field was
obtained between 1997 December 15 and 1998 January 20, with 
a total exposure time of 61,269~s.
This deeper observation detected a number of fainter X-ray sources 
above a limiting unabsorbed flux of 
$\approx 2 \times 10^{-14}$~erg~cm$^{-2}$~s$^{-1}$,
including four of the five previous sources, as well as
10 new ones.  Nine sources fall within the 99\% confidence ellipse of \source.
All of these sources are listed in Table~1,
together with information about their optical identifications, which
are radio-quiet QSOs or coronal emitting stars.
The HRI astrometry was recalibrated using the
optical counterparts of five well-localized X-ray sources, for
which an average translation of $2.\!^{\prime\prime}3$ was required.
After this shift, the five fiducial X-ray sources have a dispersion of only
$0.\!^{\prime\prime}8$ from their optical positions.  In Table~1
we list optical position, or recalibrated X-ray position
in the case that no firm optical identification has been made.  X-ray fluxes
are calculated assuming a power law of photon index
--2.0 and the full Galactic $N_{\rm H}$ for QSOs and unidentified sources,
and a Raymond-Smith thermal plasma of $T = 3 \times 10^6$~K and
$N_{\rm H} = 1 \times 10^{20}$~cm$^{-2}$ for stars.

An \asca\ observation took place from 1998 April 20--22 for a total
clean exposure time of $68,900$~s in each of the two
Gas Imaging Spectrometers (GIS).  Figure~1 shows the combined GIS image. 
The detection threshold for this \asca\ observation
was $1.1 \times 10^{-13}$~erg~cm$^{-2}$~s$^{-1}$ (1--10~keV)
assuming a photon index of --1.7.  
Several sources are
detected far from the EGRET error ellipse, and only one faint
source falls within it, a radio-quiet QSO at $z = 0.973$ that was
also detected by \ro.
In Table~1 we give information about this and four additional
\asca\ sources outside the EGRET error ellipse that we were able to identify.

Diffuse X-ray emission at the western
edge of the \asca\ GIS image appears to be coming from an uncatalogued cluster
of galaxies that is evident on our CCD images.  We have not attempted
to measure the X-ray flux of this source as it is too close to the edge
of the detector and may extend outside it.
The brightest galaxies
in this vicinity are members of the cluster at $z=0.102$ and
have $R \approx 14$ and $R \approx 15$, at J2000 coordinates
$18^{\rm h} 32^{\rm m} 38.\!^{\rm s}01,
+59^{\circ} 23^{\prime} 43.\!^{\prime\prime}8$,
and $18^{\rm h} 32^{\rm m} 49.\!^{\rm s}52,
+59^{\circ} 21^{\prime} 49.\!^{\prime\prime}4$, respectively.
This X-ray source is well outside the \source\ error ellipse,
and we have no reason to suspect that they are related.  In particular,
there is no evidence of an AGN in this cluster.

The field of view of the
\asca\ Solid-state Imaging Spectrometer (SIS) detectors,
even when operated in 4-CCD mode during this
observation, is too small to cover the EGRET error ellipse.  No X-ray
sources were detected in the SIS images, so we do not discuss them
further here.

\subsection{Radio Observations}

We reduced an archival VLA observation of this field which
was taken at a frequency of 1.4 GHz on 1995 February 21 
in the D configuration.
We found 14 sources stronger than 2.5~mJy in the neighborhood
of \source.  They have a positional accuracy
of approximately $7^{\prime\prime}$ for the fainter sources,
and $1^{\prime\prime}$ for 
sources stronger than 15~mJy.  For completeness, we examined
the NVSS catalog at the same frequency to confirm six 
more faint sources that were 
marginally detected in the 1995 pointing.
To incorporate information at other radio frequencies, we searched
the Westerbork Northern Sky Survey (WENSS), which covered this field
to a limiting flux of 18mJy at 326 MHz (Rengelink et al. 1997),
and the NRAO 4.85 GHz catalog of Becker \etal~(1991),
which has a flux limit of 20~mJy at this location.
A combined total of 20 radio sources were found
inside and outside the error ellipse.  Their properties
are listed in Table~2, and their positions are shown in Figure~2.
Most notably, there are no flat-spectrum sources in this field,
and there are only three sources 
within the 99\% confidence error ellipse of \source, all fainter
than 4~mJy at 1.4~GHz.

\section{Optical Spectroscopy and Results}

We used a number of spectrographs to obtain moderate-resolution spectra
of candidate X-ray and radio counterparts as well as UV excess objects
selected from our optical imaging survey.  These instruments include the 
Goldcam spectrograph on the KPNO 2.1m telescope, the 
Mark III spectrograph on the MDM 1.3m McGraw-Hill and 2.4m Hiltner telescopes,
the Kast double spectrograph on the 3m Shane reflector at Lick Observatory,
the Low  Resolution Spectrograph (LRS) on the Hobby-Eberly telescope,
and the Low Resolution Imaging Spectrograph (LRIS) on the Keck II telescope.
Most spectra were analyzed independently by two authors and an agreement on
classification was reached after comparing separate findings. The 
spectra were analyzed for emission and absorption lines and classified 
as either as star, galaxy, white dwarf, AGN, or uncertain.
We have completed spectroscopy to a limiting magnitude of $B = 20.3$,
which includes 43 out of 53 optical candidates.
In addition we have spectra of two objects 
fainter than $B = 20.3$.  Finding charts for the classified 
objects are given in Figures~3 and 4, and their spectra are shown
in Figures~5 and 6.

Thus far we have found eight QSOs by the UV excess technique
in the magnitude range $18.5 < B < 21.3$.
Their redshifts range from 0.504 to 2.21.
These are listed in Table~3 and their positions are shown in Figure~2.
By design, they
all fall within or very close to the \source\ error ellipse.
The efficiency of our color selection agrees fairly well with the
number counts reported by Koo \& Kron (1998) and Hall \etal~(1996), which
would predict that six QSOs with $B < 20.3$ and $z < 2.3$
would be found within a region of this size.  Several additional
candidates were found to have featureless blue spectra
that we cannot securely classify.  Since their
colors are consistent with those of white dwarfs, we suspect
that they are of the weak-lined (DC) variety.

Of the X-ray sources, six have been identified with radio-quiet QSOs,
including five that were independently selected by UV excess colors.
A seventh X-ray quasar is an \asca\ and radio source 
at $z = 0.668$ 
that lies well outside the EGRET error circle.  Four more X-ray sources are
identified with coronal emitting stars of types G, K, and dMe whose
X-ray fluxes are normal for their optical magnitudes.  

Two radio sources outside the EGRET error ellipse are identified
with bright, early type galaxies at redshifts of 0.106 and 0.156, respectively,
that lack any emission lines or evidence of non-stellar continuum in
their optical spectra (Figure~6).  Neither of these are promising 
$\gamma$-ray source candidates.   The lower-redshift galaxy is 
close to the X-ray emitting galaxy cluster that is west 
of the EGRET error ellipse 
and it is apparently a member of the cluster.

We have had less success in identifying the faint radio
sources within the EGRET error ellipse.  Bright optical objects near their
positions have proven to be ordinary stars, indicating that their
true optical counterparts are likely to be fainter than our limiting
magnitude for spectroscopy.  Finding charts for both of the radio
galaxies, as well as for several unidentified radio sources, are
displayed in Figures~3 and 4. 

\section{Notes on Individual Interesting Objects}

RX~J1834.1+5913: This is the brightest quasar in the EGRET error
ellipse
($V=18.8,\ z = 0.973$) and it is detected by both \asca\ and \ro.
Its X-ray flux decreased between the two
ROSAT observations, from $1.9 \times 10^{-13}$~erg~cm$^{-2}$~s$^{-1}$
in 1995 to $4.76 \times 10^{-14}$~erg~cm$^{-2}$~s$^{-1}$ in 1997--98.
However, we are cautious about this variability since the source was
near the edge of the detector in the later observation. 
We have several optical measurements of it in 1997, 1998,
1999 which also show modest variability.  The largest change of
$0.39$ magnitudes occurred between 1998 June and 1999 September,
but there is no evidence for rapid variability on time scales
of days.  In addition, the equivalent width of its Mg~II emission line
did not vary in spectra taken at two different epochs.  Thus, the spectral
and variability properties of RX~J1834.1+5913 offer no strong reason to
argue that it is a candidate identification for \source.  However,
as the brightest QSO in the EGRET error ellipse, it does warrant
continued scrutiny.  In \S 5, we compare the properties of this source
to those of the identified EGRET blazars in order to illustrate how unusual
any AGN counterpart of \source\ must be.

UVQ J1834.3+5926: At $z = 2.21$, this is the highest redshift QSO that
we have found near \source.  Its optical spectrum is somewhat
unusual in that it is the reddest of all the QSOs in this field,
and its emission lines are broad but weak.  We suspect that its
Ly~$\alpha$ line, which falls just blueward of our Keck spectrum,
is responsible for boosting its $U$-band flux and helping it to meet
the UV excess criterion.

RX~J1834.4+5920: This relatively bright \ro\ source
($5.3 \times 10^{-14}$~erg~cm$^{-2}$~s$^{-1}$, assuming a
$T = 3 \times 10^6$~K thermal plasma spectrum) remains unidentified,
although it lies near the edge of the HRI detector where the point-spread
function is very poor.  An M star of magnitude $R = 17.8$ has been
suggested as a possible identification even though it lies $15^{\prime\prime}$
from the X-ray position (Carrami\~nana \etal~2000). 

VLA J1834.7+5918:  This faint radio source of 3.7~mJy remains without 
spectroscopy, yet a blue optical object with $V=21.4$ falls just inside
the western boundary of its error circle (see Figure~3).
Although lacking X-ray emission,
it is still a possible quasar or BL~Lac object and worth further study,
especially spectroscopy of the optical candidate.  Since this is the brightest
and most promising radio source of those within the EGRET error ellipse,
we adopt its radio flux as an upper limit for \source\ in subsequent
discussion.

VLA J1835.1+5906:  This is the brightest radio galaxy 
($R=15.1,\ z = 0.156$) at the edge of the
EGRET error ellipse.  Its optical spectrum was examined for any evidence 
of a BL~Lac object in its nucleus, the principal indicator of which would be
a shallower than normal break at 4000 \AA.  However, no such evidence is
seen. This plus its steep radio spectrum,
$\alpha = -0.53$ between 1.4 and 4.85~GHz
and absence of X-ray emission argue
against VLA J1835.1+5906 being a BL~Lac identification of \source.

VLA J1835.6+5939 (=AX~J1835.7+5939):  This is a quasar at $z = 0.668$
and the brightest radio source near \source,
with a 1.4 GHz flux of 359 mJy.  However, it is outside
of the 99\% error ellipse by $8^{\prime}$,
and this plus its steep radio spectrum,
$\alpha = -0.84$ between 1.4 and 4.85~GHz,
argue against considering it as a strong $\gamma$-ray candidate.

RX~J1836.2+5925: This is perhaps the most intriguing object found in
all of our searches.  It was the brightest X-ray source
within the error ellipse ($1.6 \times 10^{-13}$~erg~cm$^{-2}$~s$^{-1}$),
at least during the second \ro\ observation, but it
was undetected in the first \ro\ pointing or in the \asca\ observation.
Thus, it must have varied by at least a factor of 2 in the long
term, although it emitted steadily over the one-month span which 
comprises the second \ro\ observation.
RX~J1836.2+5925 of interest here primarily because it does not have 
an optical counterpart in any color (Figure~7) to limits of
$U>22.3, B>23.4, V>23.3$, and $R>22.5$.  
A red stellar object of $R \approx 19.7$ is located
$11.\!^{\prime\prime}4$ west of the
X-ray centroid, but it is not a viable candidate given the
precision with which several other X-ray sources in this
field line up with their established optical counterparts.  As described
above, the HRI astrometry in this figure was recalibrated using the
optical counterparts of five well-localized X-ray sources, for
which an average translation of $2.\!^{\prime\prime}3$ was required.
After this shift, the five X-ray sources have a dispersion of only
$0.\!^{\prime\prime}8$ from their optical positions.  Thus, the
illustrated error box which is $8^{\prime\prime}$ on a side must
include the true position beyond a reasonable doubt.

None of the optical objects near the error box of RX~J1836.2+5925
show any proper motion which could account
for their positional discrepancy with the X-ray source.
We have not obtained spectroscopy for any of these faint neighbors,
but a deeper and more exhaustive optical study of this X-ray source
would be important to evaluate its qualifications as a possible new type
of $\gamma$-ray source counterpart.  By the definition of Stocke et al. (1991),
this X-ray source has an X-ray to optical flux ratio $f_X/f_V > 78$.
Such a high ratio is found only among low-mass X-ray binaries and 
isolated neutron stars.  As we argue below, neither 
of these object classifications would make \source\ compatible with the 
broad-band spectra of the well-identified EGRET sources.

If RX~J1836.2+5925 is not the counterpart of \source, then it might
be similar to the newly discovered class of luminous soft X-ray
transients that have been found by \ro\ in the nuclei of non-active
galaxies.  These are as luminous as $10^{44}$~erg~s$^{-1}$ and last
for several months (Grupe, Thomas, \& Leighly 1999;
Komossa \& Greiner 1999; Komossa \& Bade 1999). A promising interpretation
of these events is
tidal disruption and accretion of stellar debris by a central black
hole.  If RX~J1836.2+5925 is such an event, then it could reside in a host
galaxy at $z \approx 0.5$ which deeper optical imaging could detect.

\section{Multiwavelength Comparisons to Known Classes of EGRET Sources}

\subsection{Blazars}

Our radio, optical, and X-ray data on active objects
in the field of \source\ can be compared with
other identified EGRET sources to evaluate whether \source\
can still fall within the multiwavelength parameters 
of any of the known classes
of $\gamma$-ray emitters.  Beginning with blazars, Figure~8
shows radio, optical, X-ray, and $\gamma$-ray fluxes of
the sample of well-identified EGRET blazars defined by
Mattox \etal~(1997, and personal communication).  \ro\ and {\it Einstein} 
fluxes are taken from
Fossati \etal~(1998), and
$V$ magnitudes and total 4.85~GHz radio fluxes from 
Mattox (personal communication).
The EGRET spectral points from \source\ are taken from Reimer \etal~(2000).
Of the numerous candidate identifications which we could superpose,
we chose two, namely, the brightest QSO within the error ellipse
(RX~J1834.1+5913, $z = 0.973$), and the brightest radio source within the error
ellipse (VLA J1834.7+5918).  For the latter, we hypothesize that the
suggestive $V=21.4$ optical identification is correct, and we graph
an X-ray upper limit from the deeper \ro\ observation.  For the QSO,
we assign an upper limit of 0.5 mJy at 1.4~GHz, from the VLA image.

The smooth curves fitted to these two candidates
correspond to the sum of two empirical third-order polynomials
as applied by Comastri \etal~(1995).  This is not a model
of blazar emission, but only a guide to the eye in making
empirical estimates of the peak fluxes
at low and high energy.  In doing so we assume the presence of two 
emission mechanisms, a low-energy synchrotron component
and a high-energy component peaking in the $\gamma$-ray band,
possibly due to inverse Compton scattering.

While the optical and X-ray properties of our brightest candidates
are not unprecedented, they lie at the faint end of the distributions.
In particular, the X-ray upper limit for VLA J1834.7+5918
(or any of the other radio sources in the error ellipse) falls
below the faintest blazars by at least an order of magnitude.
More significant are their faint radio fluxes which, in the case of
VLA J1834.7+5918 is two orders of magnitude fainter 
than the faintest radio counterpart of any well-identified EGRET
blazar.  RX 1834+5913 is nearly three orders of magnitude fainter in the
radio band.
Figure~9, in which
the ratio of
4.85~GHz flux density 
to the peak $\gamma$-ray flux in the range $E>100$~MeV
is graphed as a function of $\gamma$-ray flux
for the Mattox blazars, confirms the
highly discrepant positions of any of the QSOs or radio sources
which are positionally coincident with \source\ and candidates for
identification with it.
(We assume in this Figure a flat radio spectrum for \source,
since none of its faint candidates were actually detected at 4.85~GHz.)

Another property of the majority of EGRET blazars is their rapid and
large-amplitude flux variations.  The absence of such obvious 
$\gamma$-ray variability from \source\ already argues against a blazar
nature for it (Reimer \etal~2000).
We have also searched our $V$-band images obtained in 1997 and 1998
for objects with rapid or extreme optical variability, looking for 
variations of $\Delta V > 0.3$.  Apart from the modest variability
of the $z = 0.973$ QSO RX~J1834.1+5913 described above,
no optical candidates for blazar activity were discovered in this manner.

\subsection{Rotation-Powered Pulsars}

Similar to the comparison with known blazars, we can examine
how \source\ compares to the EGRET pulsars.  In Figure~10, we
compare the 0.1--2.4 keV X-ray flux (Becker \& Tr\"umper 1997)
and average flux $E>100$ MeV for EGRET pulsars (Fierro 1995;
Kaspi \etal~2000; Ramanamurthy \etal~1995).  Any possible pulsar
counterpart of \source\ should be assigned an X-ray flux upper
limit equal to the flux of the brightest unidentified \ro\
source in the error ellipse.  This role is therefore
properly assigned to RX~J1836.2+5925, although the fact that it
is variable in X-rays already places some doubt upon its credentials
as a pulsar candidate.  Most of the soft X-ray flux observed from
intermediate-age neutron stars is surface thermal emission, which
should not vary from year to year.  However, the additional nonthermal X-ray
component which is present in Geminga and other $\gamma$-ray pulsars
could in principle vary, and Halpern \& Wang (1997a) suggested
that it does in Geminga.  Therefore, we use the quiescent flux upper
limit of this source (from the 1995 \ro\ observation)
for comparison in Figure~10.  Such a comparison
strains the analogy with Geminga.  While the latter is a cooling neutron
star with $T \approx 5.6 \times 10^5$~K at $d \geq 150$~pc,
\source\ is about 50 times fainter in X-rays, thus either $d > 1$~kpc,
or if it is to be located at a similar distance as Geminga,
its surface temperature should be less than $3 \times 10^5$~K.  
The larger distance is problematic, since it implies a $\gamma$-ray
luminosity of $1.7 \times 10^{35}\ (d/1\ {\rm kpc})^2$~erg~s$^{-1}$
if isotropic, which is at least
5 times larger than the spin-down power of Geminga,
$3.3 \times 10^{34}$~erg~s$^{-1}$.
Alternatively, if it is closer than 1~kpc, then its surface must be
cooler and it is likely to be older than $3 \times 10^5$~yr,
which would also strain its $\gamma$-ray efficiency.
If \source\ is a pulsar but RX~J1836.2+5925 is {\it not} its counterpart,
then its X-ray flux upper limit is reduced to $\approx 5 \times 10^{-14}$
erg~cm$^{-2}$~s$^{-1}$, or 80 times fainter than Geminga.

Younger pulsars such as Vela, PSR~B1951+32,
and PSR~B1706-44 are EGRET sources with luminosities in the range 
$(1-2) \times 10^{35}$~erg~s$^{-1}$, but \source\ lacks the nonthermal
X-ray emission and synchrotron nebulae that accompany those
more luminous pulsars.  Furthermore, it would be highly unexpected
to find a pulsar of characteristic age
$\tau < 1 \times 10^5$~yr at $d \sim 400$~pc from the Galactic plane.
since this would require a kick velocity
$v > 5000\ (\tau/ 10^5\ {\rm yr})^{-1}$~km~s$^{-1}$.
An interesting possibility would be a recycled millisecond pulsar,
which could be old yet energetic.
But even such pulsars manage to channel at least $5 \times 10^{-4}$
of their spin-down power into either thermal (Halpern \& Wang 1997b)
or nonthermal X-rays (Becker \& Tr\"umper 1999; Mineo et al. 2000).
In the case of \source, any pulsar counterpart would have
$L_X(0.1-2.4\ {\rm keV})/L_{\gamma}(>100\ {\rm MeV}) < 6 \times 10^{-5}$,
which places  a uniquely low limit on the ratio of X-ray to spin-down power.

\subsection{Other Possible $\gamma$-ray Sources}

In addition to the well established classes of $\gamma$-ray blazars and
pulsars, several associations have been suggested which are highly plausible
even while not conclusively proven.  Most notable is the radio star
and Be/X-ray binary LSI~$+61^{\circ}303$ (Strickman et al. 1998), long
associated with the $\gamma$-ray source 2CG~135+01.  Similar objects
might be the 47~ms pulsar B1259--63 with a Be star companion, detected
up to 200~keV (Grove et al. 1995),
and the Be/X-ray binary SAX~J0635+0533 in the error circle 
of 2EG J0635+0521 (Kaaret et al. 1999).  Since these systems all have neutron
stars with Be star companions, their $\gamma$-ray emission is
not necessarily confined to the pulsar magnetospheric
mechanism, but may arise
in the interaction of the relativistic pulsar wind with the wind of the
companion, or with its radiation.  However, all of these systems have
bright optical companions, as well as strong X-ray emission at least at
some of the time.  It is estimated that there are only 200 Be/X-ray binaries
within 5~kpc (Rappaport \& van den Heuvel 1982); these are young systems
which are confined to the Galactic disk.  If a Be star binary,
the location of \source\ well away from the Galactic plane
would probably make it the nearest such system, and virtually impossible
to miss since its $V$ magnitude would be brighter than 9 if at $d < 1$~kpc.
Since no such Be star is present in this region,
this scenario for \source\ can safely be ruled out. 

\section{Implications for EGRET Source Identifications}   

The statistical issues concerning the identification of 
EGRET sources with flat-spectrum radio sources were 
rigorously addressed by Mattox et al. (1997), and it is hardly possible
to improve upon that analysis at this time.  To summarize,
flat-spectrum radio
sources are the only AGNs that have been detected by EGRET
with any degree of confidence.  Unfortunately, while the
EGRET survey is flux limited, the radio
identifications of EGRET sources in Figure~9 are {\it not} flux limited,
but rather are plagued by source confusion due to the
large size of the EGRET error circles and the large surface
density of radio sources.  Thus, the statistical reliability of EGRET
source identifications is lower than that in any other branch of
astronomy.  As Mattox et al. calculate, the radio sources that are reasonably
secure ({\it i.e.}, $> 95\%$ confidence) identifications of
EGRET sources have 5~GHz flux densities $ > 500$~mJy.
That is why the correlation between radio flux and $\gamma$-ray flux
in Figure~9 is weak and less than linear.
Below 50~mJy, it is not even possible to make a meaningful argument for
identification because the mean separation of such radio sources on
the sky is comparable to the size of the EGRET error circles.
Accordingly, there are {\it three} radio sources within the error circle
of 3EG~J1835+5918, and they are all fainter than 4~mJy.  None is an X-ray
source.
If any one of these radio sources were the true counterpart
of the EGRET source, its
ratio of radio to $\gamma$-ray flux would be two orders
of magnitude smaller than that
of any known blazar.  We are not claiming that 3EG~J1835+5918 is unique in
this regard.  Other unidentified EGRET sources may eventually prove to
be similar.

Even though we cannot yet point to a likely identification of \source,
it is apparent from our multiwavelength observations that the true counterpart
must be physically different or extreme in its properties
relative to the classes of EGRET sources that have been identified so far.
This is true whether the counterpart is one of the candidates studied here,
or an undetected fainter object.  Furthermore, it is unlikely
that a systematic error in $\gamma$-ray position has caused us to
overlook a more conventional identification.  In radio and X-ray we have
explored
a region approximately 4 times the size of the 99\% confidence location,
and even within this ample area there are no blazar or pulsar candidates.
For example, even if the counterpart were the brightest radio quasar in
Figure~1, which is $8^{\prime}$ from the edge of the 99\% confidence region,
that object is a steep-spectrum radio source, as are all of the 
other bright radio sources outside the EGRET error ellipse.  Thus,
an error of this magnitude in the location of \source\
will not change the basic conclusion that a new or extreme type
of counterpart is responsible.
 
One possible implication of this result is that radio-steep or radio-quiet
quasars could be counterparts of some of the unidentified EGRET
sources, despite the analysis of Mattox \etal~(1997) which argues that
such a new population is not needed.
Instead of interpreting the hard $\gamma$-ray spectrum and
lack of variability as pulsar-like, it might be that these properties are
also characteristic of the less violently variable AGNs.  The obstacle
to identifying a potential radio-weak or radio-quiet EGRET source population 
is not sensitivity, but source overlap.  There are simply too many such
AGNs in any EGRET error circle.  While it is almost 
certainly the case that
weaker radio blazars will be identified with high-energy $\gamma$-ray sources
once their error circles are reduced by {\it GLAST\/}, it remains to be
seen whether or not qualitatively different types of AGN will be
also be represented.

An interesting scenario for a new type of $\gamma$-ray AGN has
been suggested by Ghisellini (1999), who posits the existence of
blazars whose synchrotron spectrum peaks in the MeV band,
and an inverse-Compton component that peaks in the TeV.
A variation of such a model could fit the 
multiwavelength spectrum of the $z = 0.973$ QSO RX~J1834.1+5913
or any of the fainter QSOs in the field provided that 
the proper index for the power-law electron energy
distribution can be accommodated,
and only if the observed optical emission is dominated by the usual
thermal accretion-disk emission so that it can represent an upper
limit to the underlying synchrotron power law.   In such a model
the hard X-ray emission is due entirely to the synchrotron component.
The absence of a radio counterpart is naturally explained by the
form of the power law, which in this case
requires a flat spectral index 
$\alpha \approx -0.45$ where $F_{\nu} \propto \nu^{\alpha}$,
thus
the power-law index of the electron energy distribution is $p \approx  -1.9$ .
Such a prediction can easily be tested by more sensitive
hard X-ray spectra of the QSO RX~J1834.1+5913.

Radio-quiet blazars have been hypothesized theoretically
(Ghisellini 1999; Mannheim 1993; Schlickeiser 1984)
but so far none have been identified (Stocke \etal~1990; 
Jannuzi \etal~1993), and it is not even clear what such
a phenomenon would mean.
Could the multiwavelength properties of \source\ be
evidence of the hadronic model, the so called proton
blazar?  Such a theory proposes
to explain $\gamma$-ray emission in blazars, relying on protons 
accelerated by shocks moving through the jet. The accelerated protons
then interact with soft-photons which lead to the creation of pions that
further decay and cascade into electron-positron pairs, $\gamma$-rays and
neutrinos. Such a model (Mannheim 1993) could fit the observations of
\source\ if the energy density ratio of protons to electrons is greater than 10.
    
If \source\ is a pulsar, it implies that highly efficient
(or highly beamed) $\gamma$-ray pulsars can avoid producing 
soft X-rays at a level below $10^{-4}$ of their apparent
$\gamma$-ray luminosity.  At least two mechanisms of X-ray
emission have been observed to accompany all $\gamma$-ray
pulsars at such levels or higher (Wang et al. 1998).
In the outer-gap model,
synchrotron emission from secondary pairs 
that are produced by conversion of $\gamma$-rays in the
inner magnetosphere where $B > 2 \times 10^{10}$~G
can explain the nonthermal X-ray component from pulsars
like Geminga and PSR~B1055--52.  The second mechanism
is thermal emission arising from the heated polar caps that
are impacted by the inward-going accelerated particles
from the outer-gap accelerator.  There
is good evidence that polar-cap heating
occurs even in recycled pulsars which are {\it not}
detectable EGRET sources (Zavlin \& Pavlov 1998; Halpern \& Wang 1997b).
Therefore, it is difficult to reconcile such a theory, as well as the
observational fact that pulsars are X-ray sources of
$L_X > 10^{-4}\ I\Omega\dot\Omega$, with a pulsar origin for \source.
If many of the unidentified EGRET sources are similar
radio-quiet pulsars in the Galactic plane, X-ray absorption
makes them exceedingly difficult to identify, and perhaps they will
be revealed only when $\gamma$-ray observations are sensitive
enough to detect their pulsations independently.

\section{Conclusions and Further Work}

We identified all but one of the 
X-ray sources in the field of \source\ to a flux limit of approximately
$5 \times 10^{-14}$~erg~cm$^{-2}$~s$^{-1}$.
These are radio-quiet QSOs [$F(1.4\ {\rm GHz}) < 0.5$~mJy],
coronal emitting stars, and a cluster of galaxies.
There are no flat-spectrum radio sources 
in the vicinity to a flux limit of $\approx 20$~mJy,
and no radio sources in the EGRET error ellipse brighter than
4~mJy at 1.4~GHz.
In addition, we find no evidence of
a BL~Lac object hosted in any low-redshift galaxy.
We also found several QSOs, as one would expect,
using purely optical color selection.  Multiple-epoch
optical imaging of the entire EGRET error ellipse has not revealed
any notable variability.
The discovery of only radio-quiet quasars in the error circle 
of \source\ is a sobering development in the search for its identification.
Although the $\gamma$-ray properties of \source\ are more similar to
those of Geminga and other EGRET pulsars, no other indirect evidence for
a pulsar, apart from one unidentified X-ray source (RX~J1836.2+5925) whose
optical counterpart is probably fainter than $B=23.4, V=23.3$,
and $R=22.5$, has been found.  Yet, the fact that this X-ray source
is variable by at least a factor of 2 would make it unique among
rotation-powered pulsars.
Taken together, these findings point to the possibility of a truly 
remarkable object, one that cannot be matched by any 
known class of $\gamma$-ray source.

Even in the absence of a definite identification, it is clear
that \source\ is lacking in one or more of the physically
essential attributes of any known class of $\gamma$-ray emitter.
Its radio flux is at least two orders of magnitude fainter
than any of the securely identified EGRET blazars,
and its soft X-ray flux is at least 50 times
fainter than that of Geminga and similar EGRET pulsars.  If it is an AGN
it lacks the beamed radio emission of blazars.  If it is an isolated
neutron star, it lacks the steady thermal X-rays from a cooling surface
and the magnetospheric non-thermal X-ray emission that is characteristic
of all EGRET pulsars.  If a pulsar, \source\
must be either older or more distant than Geminga,
and probably an even more efficient or highly beamed $\gamma$-ray engine.

We have plans to complete the optical spectroscopy of fainter
candidates in this field to $B \approx 21.5$ and we will also
study fundamental properties such as polarization and optical
variability of the newly discovered AGNs.  Perhaps the most
important technique which we have not yet applied is polarimetry.
Polarimetry provides a definitive test for synchrotron emission
in an ordered magnetic field, and polarization is one of the essential
properties of blazars.  Perhaps the blazar nature of a radio-quiet
beam in an AGN can only be demonstrated in this way.  A deeper
radio pulsar search would also be warranted.
Finally, we will pursue the optical identification of the \ro\ source
RX~J1836.2+5925 to the faintest magnitudes that are necessary in order
to find our whether or not it is a neutron star.
In combination, these observations may result in the identification
of an important EGRET source, and possibly the prototype of a new class of
$\gamma$-ray emitter.

\acknowledgments{ }
We thank Eric Gotthelf for his assistance with the reduction
of \asca\ data, Karen Leighly and John Tomsick for assistance
with the optical imaging, and John Mattox,
Greg Madejski, Brenda Dingus, and Reshmi Mukherjee for helpful
discussions. 
This work was supported by NASA grants NAG 5-3229 and NAG 5-7814.

\clearpage

\begin{deluxetable}{cccclcc}
\tablenum{1}
\tablecaption{X-ray Sources in the Field of \source\  }
\tablehead
{
Name & R.A. & Decl. & $B$ & \omit \hfil $z$ \hfil & $F_X(0.1-2.4\,{\rm keV})$\,\tablenotemark{a}
& ID \\
& (h \hskip 0.5em m \hskip 0.5em s) & ($\circ\ \ \prime\ \ \prime\prime$) &&& 
(erg~cm$^{-2}$~s$^{-1}$)
}
\startdata
AX~J1832.6+5923\,\tablenotemark{b} & 18 32 38.00
& 59 23 43.8    & -- & 0.102 & -- & Cluster\\
AX~J1833.3+5928\,\tablenotemark{b} & 18 33 19.62
& 59 30 05.7    & 18.9 & 0.942 & $6.7 \times 10^{-13}$ & QSO\\
RX~J1834.1+5913 & 18 34 08.24
& 59 13 51.0    & 19.3 & 0.973 & $1.9 \times 10^{-13}$   & QSO\\
RX~J1834.2+5920 & 18 34 14.88
& 59 20 24.5    & 10.9 & --    & $5.3 \times 10^{-14}$   & G7 star\\
RX~J1834.3+5909\,\tablenotemark{b} & 18 34 20.36
& 59 09 15.0    & 17.8 & --    & $4.8 \times 10^{-14}$   & dMe star\\
RX~J1834.4+5920 & 18 34 24.74
& 59 20 55.6    & --   & --    & $5.3 \times 10^{-14}$   &
M5 star?\,\tablenotemark{d}\\
RX~J1835.5+5915 & 18 35 32.73
& 59 15 41.1    & 16.7 & --    & $1.1 \times 10^{-13}$   & dMe star\\
AX~J1835.6+5939\,\tablenotemark{b}  & 18 35 39.87 
& 59 39 50.7    & 18.4 & 0.668 & $2.3 \times 10^{-13}\,\tablenotemark{c}$ & RL Quasar\\
RX~J1835.9+5923 & 18 35 53.71
& 59 23 29.6    & 19.6 & 1.87  & $4.3 \times 10^{-14}$   & QSO\\
RX~J1835.9+5926 & 18 35 58.49
& 59 26 17.5    & --   & --    & $5.5 \times 10^{-14}$   & --\\
RX~J1836.0+5924 & 18 36 00.36
& 59 24 53.2    & --   & --    & $2.4 \times 10^{-14}$   & --\\
RX~J1836.1+5925 & 18 36 08.03
& 59 25 05.4    & --   & --    & $2.1 \times 10^{-14}$   & --\\
RX~J1836.2+5925 & 18 36 13.82
& 59 25 28.9    & --   & --    & $1.6 \times 10^{-13}$   & --\\
RX~J1836.6+5920\,\tablenotemark{b} & 18 36 36.90
& 59 20 41.9    & 21.3 & 1.36  & $4.1 \times 10^{-14}$   & QSO\\
RX~J1836.6+5925\,\tablenotemark{b} & 18 36 38.62 
& 59 25 25.5    & 19.8 & 1.75  & $4.8 \times 10^{-14}$   & QSO\\
RX~J1836.8+5910\,\tablenotemark{b} & 18 36 51.07 
& 59 10 08.3    & 12.1 & --    & $6.6 \times 10^{-14}$   &
K5 star\,\tablenotemark{d}\\
RX~J1837.0+5934\,\tablenotemark{b}   & 18 37 00.56
& 59 34 17.7    & 18.5   & 1.278?    & $4.2 \times 10^{-13}$   & QSO\\ 
\enddata
\tablenotetext{a}{ Unabsorbed Flux.
For QSOs and unidentified sources, $N_{\rm H}$ is taken from Dickey \& Lockman 
(1990) and $\Gamma = 2.0$ is assumed. 
For stars, $N_{\rm H} = 1 \times 10^{20}$~cm$^{-2}$
and $T = 3 \times 10^6$~K are assumed.}
\tablenotetext{b}{ Outside EGRET 99\% confidence error ellipse.}
\tablenotetext{c}{ \asca\ X-ray flux given in the 1--10 keV band.}
\tablenotetext{d}{ Carrami\~nana \etal~(2000).}
\end{deluxetable}

\begin{deluxetable}{ccccccl}
\tablenum{2}
\tablecolumns{7}
\tablecaption{Radio Sources in the Field of \source\  }
\tablehead
{
 R.A. & Decl. & $R$ & $z$ & $F(1.4\ {\rm GHz})$ & $F(326\ {\rm MHz})$ &
Comment\\
(h \hskip 0.5em m \hskip 0.5em s) & ($\circ\ \ \prime\ \ \prime\prime$) &&&
(mJy) & (mJy)
}
\startdata
 18 32 12.54 & 59 18 05.3   & --   & --    &  60.0 & 168.0 &
 No ID\,\tablenotemark{a}\\
 18 32 58.72 & 59 28 01.8   & 15.4 & 0.106 &   9.5 & 27.0  &
 Galaxy\,\tablenotemark{a}\\
 18 33 24.94 & 59 05 05.7   & --   & --    &   7.8 & --    &
 No ID\,\tablenotemark{a}\\
 18 33 42.08 & 59 11 26.6   & --   & --    &  86.0 & 280   &
 No ID\,\tablenotemark{a},$F(4.85\ {\rm GHz})=38$ mJy \\ 
 18 33 43.03 & 59 36 27.8   & --   & --    &  15.0 &  21   &
 No ID\,\tablenotemark{a}\\
 18 33 48.78 & 59 20 04.2   & --   & --    &   4.3 & --    &
 No ID\,\tablenotemark{a}\\ 
 18 33 50.55 & 59 35 31.8   & --   & --    &   5.4 & --    &
NVSS, No ID\,\tablenotemark{a}\\
 18 34 12.74 & 59 32 06.1   & --   & --    &   9.0 & --    &
 No ID\,\tablenotemark{a}\\
 18 34 43.92 & 59 24 11.2   & --   & --    &   3.4 & --    &
 No ID\\
 18 34 46.19 & 59 18 28.1   & --   & --    &   3.7 & --    &
 $V = 21.4\,?$\\
 18 34 47.16 & 59 38 31.0   & --   & --    &  25.0 & 53    &
 No ID\,\tablenotemark{a}\\
 18 34 50.88 & 59 36 54.0   & --   & --    &   3.2 & --    &
NVSS, No ID\,\tablenotemark{a}\\
 18 34 51.73 & 59 08 34.7   & --   & --    &   2.6 & --    &
NVSS, No ID\\ 
 18 35 11.71 & 59 06 46.4   & 15.1 & 0.156 & 204   & 448   &
Galaxy,\, $F(4.85\ {\rm GHz})=102$ mJy \\
 18 35 33.15 & 59 04 5.2   & --   & --    & 3.7   & --    &
NVSS, No ID\,\tablenotemark{a}\\
 18 35 39.81 &  59 39 51.9  & 18.3 & 0.668 & 359   & 1329  &
Quasar\,\tablenotemark{a}, $F(4.85\ {\rm GHz})=127$ mJy\\
 18 35 42.21 & 59 03 38.8   & --  & --     &  6.4  & --    &
NVSS, No ID\,\tablenotemark{a}\\
 18 36 31.63 & 59 05 46.1   & --  & --     &  3.2  & --    &
NVSS, No ID\,\tablenotemark{a}\\
 18 37 28.66 & 59 32 29.4   & --  & --     & 62.0  & 366\ \tablenotemark{b}   &
 No ID\,\tablenotemark{a,}\\
 18 37 31.05 & 59 31 36.1   & --  & --     & 44.0  & 366\ \tablenotemark{b}   &
 No ID\,\tablenotemark{a}\\
\enddata
\tablenotetext{a}{ Outside EGRET 99\% confidence error ellipse.}
\tablenotetext{b}{ WENSS catalogued flux corresponding to the sum of 
these two VLA sources.}

\end{deluxetable}\clearpage

\begin{deluxetable}{lcccll}
\tablenum{3}
\tablecolumns{6}
\tablecaption{QSOs Selected by Ultraviolet Excess}
\tablehead
{
\omit \hfil Name \hfil & R.A. & Decl. & $B$ & \omit \hfil $z$ \hfil & 
Telescope\\
& (h \hskip 0.5em m \hskip 0.5em s) & ($\circ\ \ \prime\ \ \prime\prime$)
}
\startdata
RX~J1834.1+5913  & 18 34 08.24
& 59 13 51.0     & 19.3 & 0.973 & Lick 3m\\ 
UVQ J1834.3+5926 & 18 34 20.65 
& 59 26 50.7     & 20.5 & 2.21  & Keck II\\
UVQ J1834.3+5918 & 18 34 21.07
& 59 18 45.8     & 20.2 & 0.504 & MDM 2.4m\\
RX~J1835.9+5923  & 18 35 53.71
& 59 23 29.6     & 19.6 & 1.87  & MDM 2.4m\\
UVQ J1836.3+5929 & 18 36 16.21
& 59 29 06.4     & 20.0 & 1.33  & MDM 2.4m\\
RX~J1836.6+5920  & 18 36 36.90 
& 59 20 41.9     & 21.3 & 1.36  & HET \\
RX~J1836.6+5925  & 18 36 38.62
& 59 25 25.5     & 19.8 & 1.75  & MDM 2.4m\\
RX~J1837.0+5934  & 18 37 00.48
& 59 34 16.3    & 18.5   & 1.278?\tablenotemark{a}    & MDM 1.3m\\

\enddata
\tablenotetext{a}{ An alternate identification of the single
emission line in the spectrum of this object as H$\gamma$
would imply $z = 0.469$.}
\end{deluxetable}
\clearpage

\figcaption[]{The combined GIS images (grey scale and contours)
from the \asca\ observation of \source.
Positions of \ro\ HRI sources (from Table~1) are indicated by crosses.}

\figcaption[]{Positions of quasars (asterisks) and radio sources 
(filled circles) in the field
of \source. Radio sources have been 
drawn in proportion to their 1.4~GHz fluxes.}

\figcaption[]{Finding charts for interesting objects selected from Tables 1--3.
Each chart is $128^{\prime\prime}$ on a side.  Circles indicate the 
statistical uncertainty in position for \ro\ and VLA sources.
Arrows indicate high-confidence optical
identifications based on the spectra displayed in Figures 4 and 5.
UV excess QSOs have no error circles, since they were optically selected
from the CCD images used to make these charts.}

\figcaption[]{Continued from Figure 3.}

\figcaption[]{Spectra of new quasars in the field of \source\
obtained via UV excess, X-ray, and radio selection.}

\figcaption[]{Continued from Figure 5.
Spectra of new quasars in the field of \source.  Also shown
are the spectra of two radio galaxies, the brightest galaxy cluster member, 
and three X-ray emitting stars.}

\figcaption[]{$BVRI$ images at the location of
RX~J1836.2+5925.  Each chart is $64^{\prime\prime}$ on a side.
The best X-ray position was derived by recalibrating the astrometry
using the optical positions of well-identified sources as
described in the text.  Although an ``error box'' is drawn
$8^{\prime\prime}$ on a side, we believe that the combined statistical
and systematic error in position is no worse than $3^{\prime\prime}$
in radius. Detection limits are $U>22.3, B>23.4, V>23.3$, and $R>22.5$.}

\figcaption[]{Asterisks are the collected radio, optical, X-ray, and
$\gamma$-ray fluxes of EGRET blazars
(Hartman \etal~1999; Mattox \etal~1997; Fossati \etal~1998).
Shown for comparison are the 
EGRET spectral points from \source\ (Reimer \etal~2000).
In the absence of an obvious identification from X-ray, optical,
or radio data, we illustrate the properties of 
the brightest QSO within the error ellipse
(RX~J1834.1+5913, circles),
and the brightest radio source within the error
ellipse (VLA J1834.7+5918, triangles).
The smooth curves fitted to these two candidates
correspond to the sum of two empirical third-order polynomials.}

\figcaption[]{The ratio of
4.85~GHz flux density from Mattox \etal~(1997)
to the peak $\gamma$-ray flux in the range $E>100$~MeV
for identified EGRET blazars
is graphed as a function of $\gamma$-ray flux.
The circle and triangle represent
the same candidates as in Figure~8.
We chose in this Figure to assume a flat radio spectrum for \source,
although none of its faint candidates are actually detected at 4.85~GHz.}

\figcaption[]{Comparison of (0.1-2.4 keV) X-ray flux (Becker \& Tr\"umper 1997)
and average flux $E>100$ MeV for EGRET pulsars (Fierro 1995;
Kaspi \etal~2000; Ramanamurthy \etal~1995).  The arrow 
corresponds to the X-ray upper limit from \ro\ observations
of \source, assuming that the brightest unidentified \ro\
source in the error ellipse could be a pulsar.}

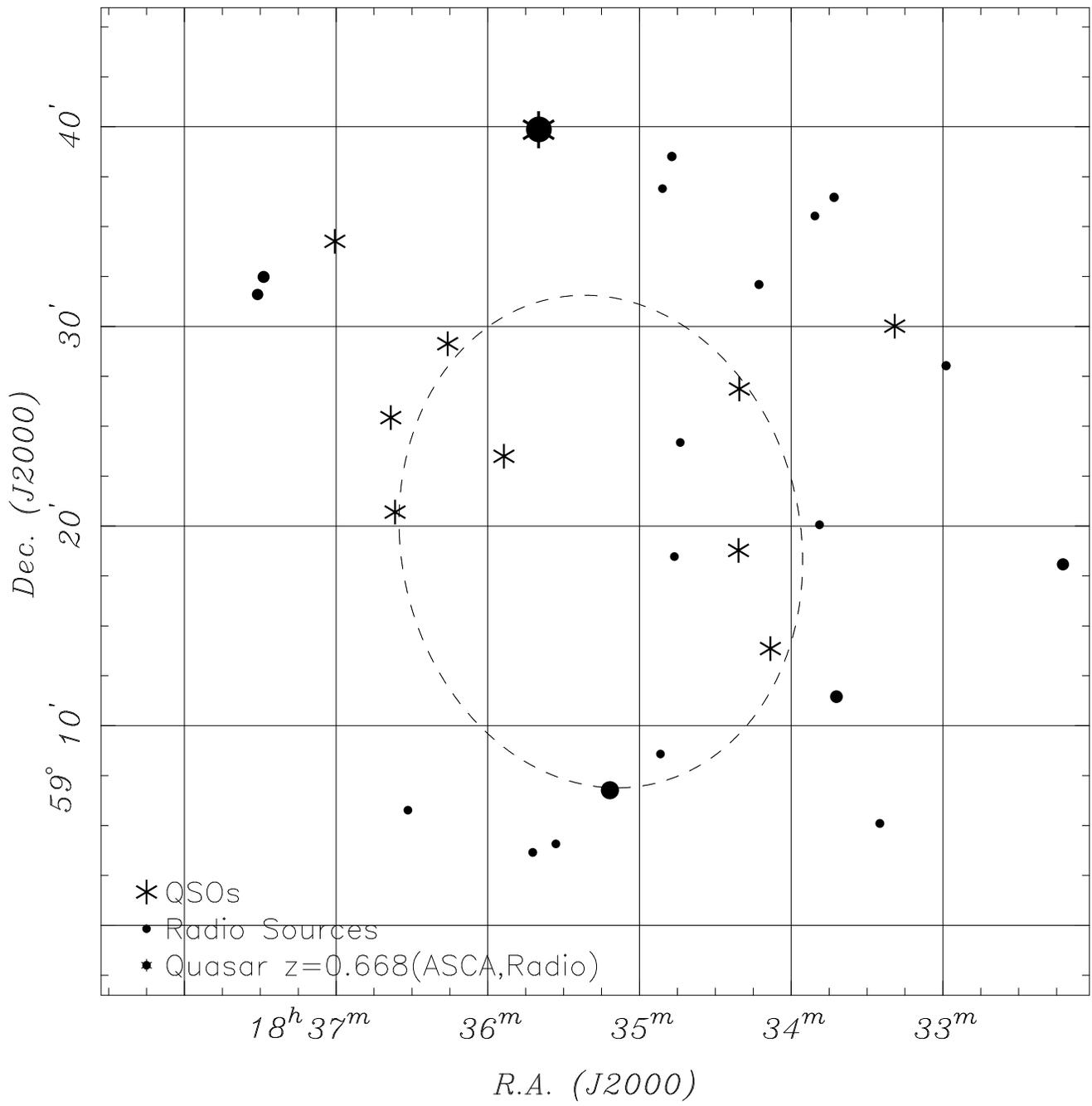
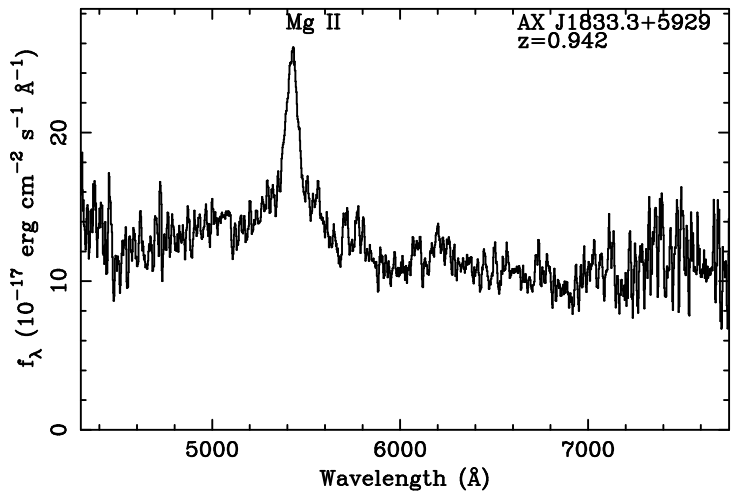
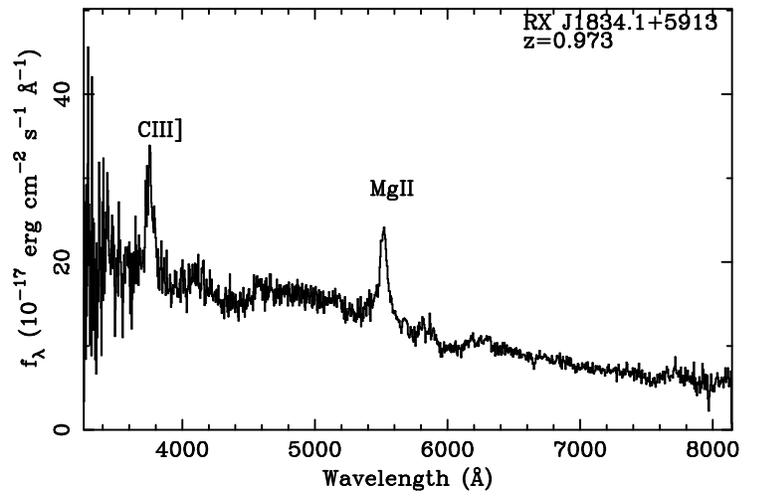
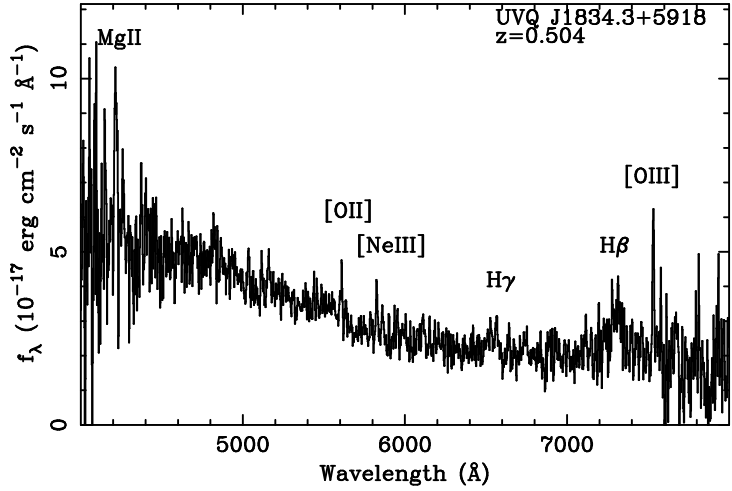
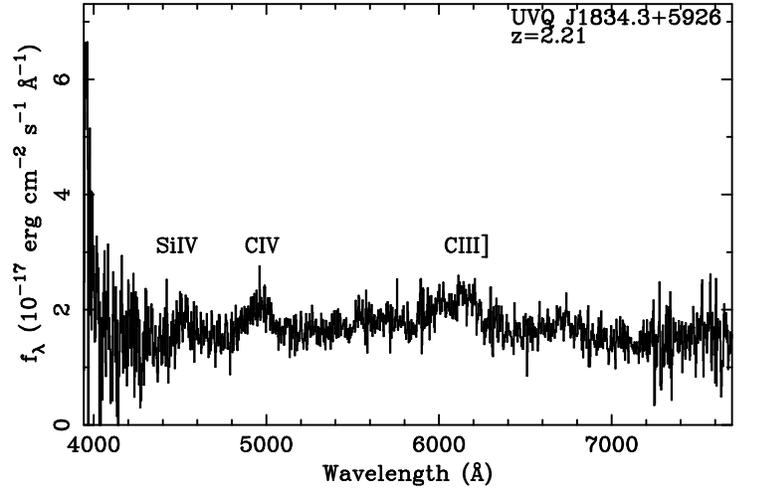
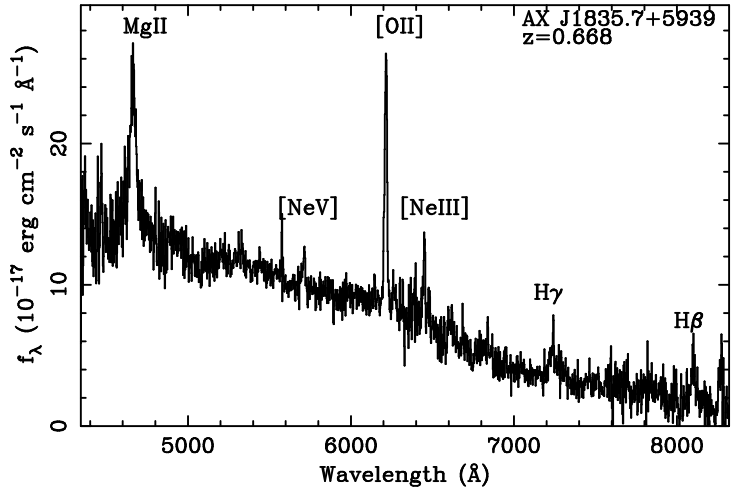
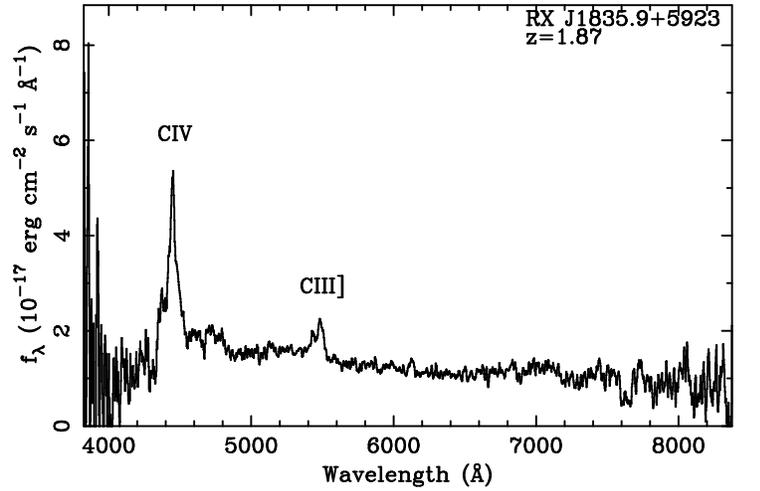
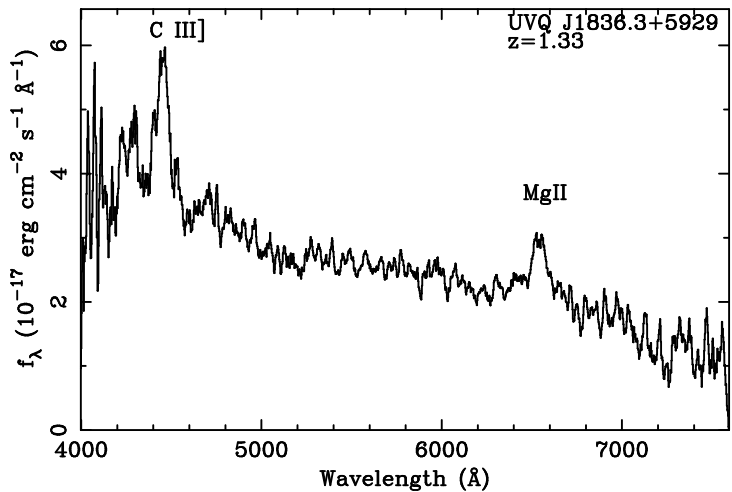
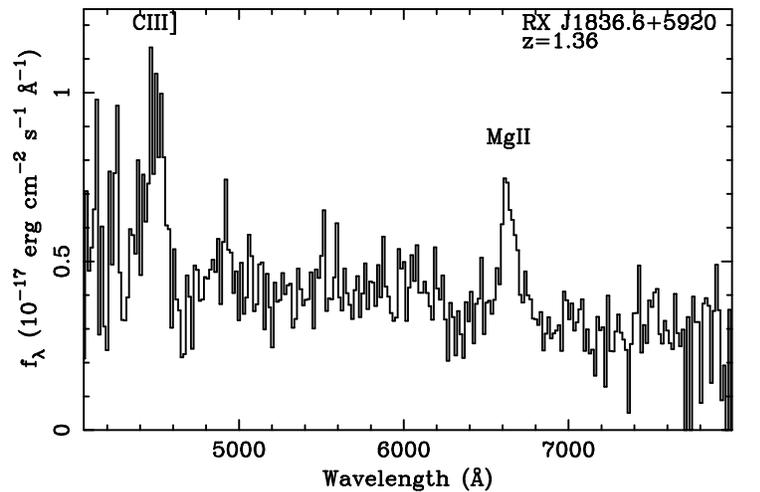
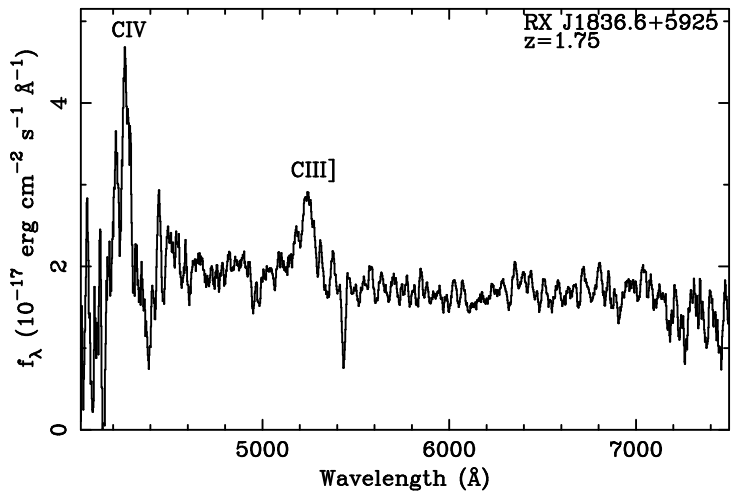
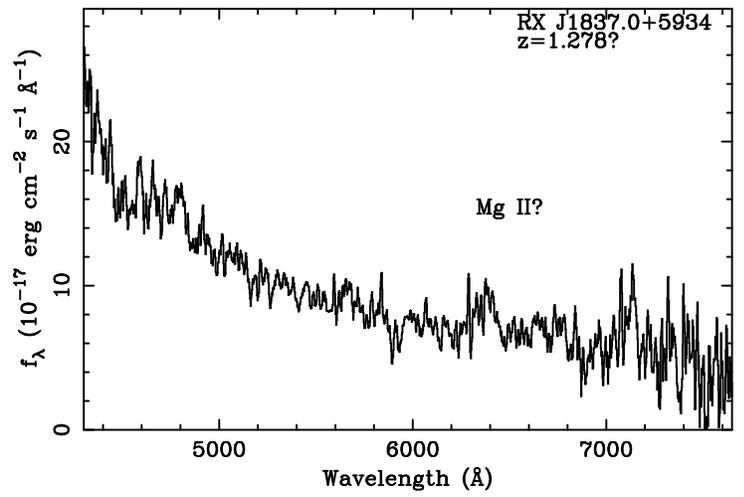
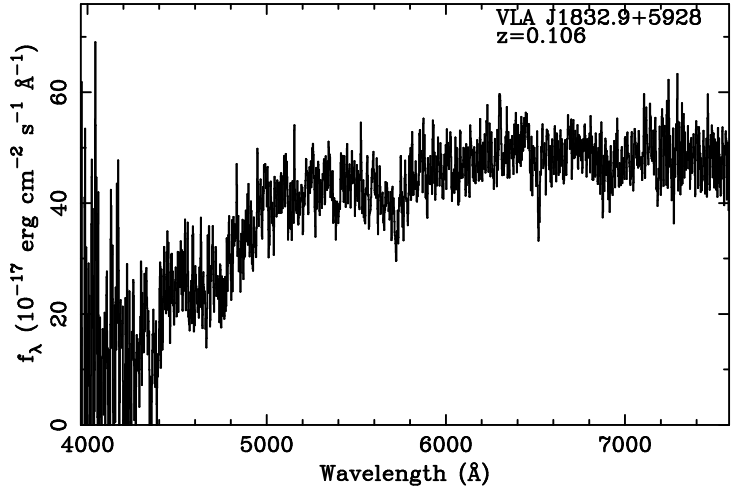
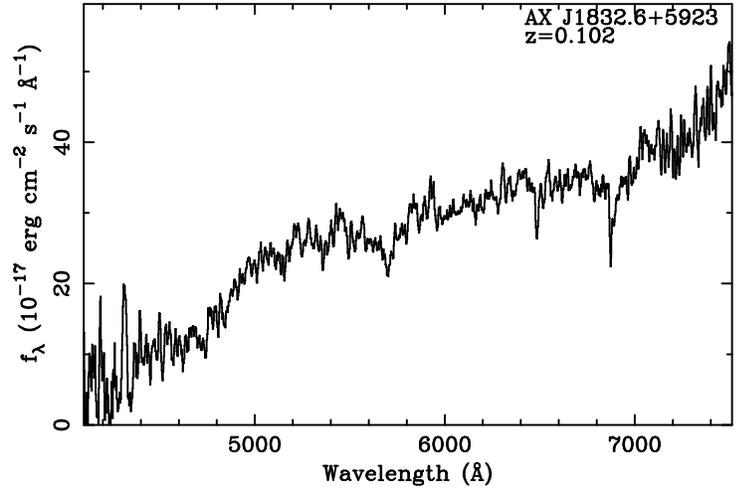
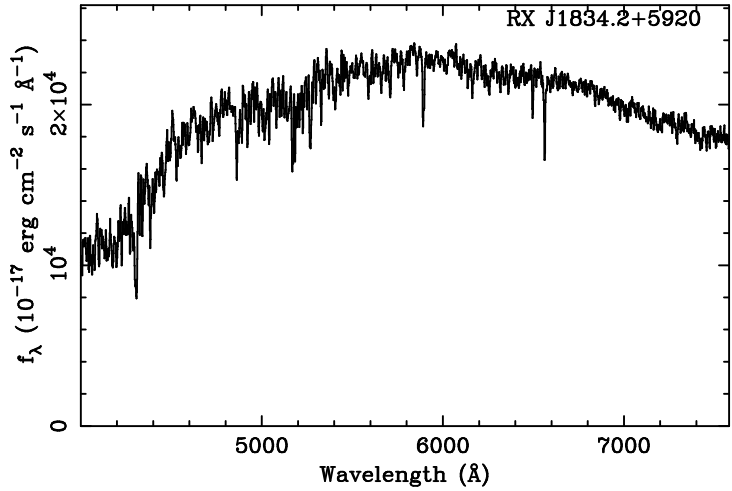
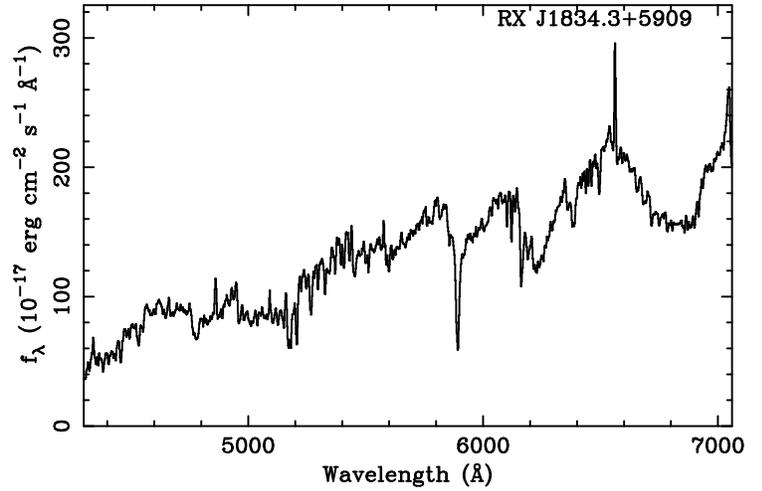
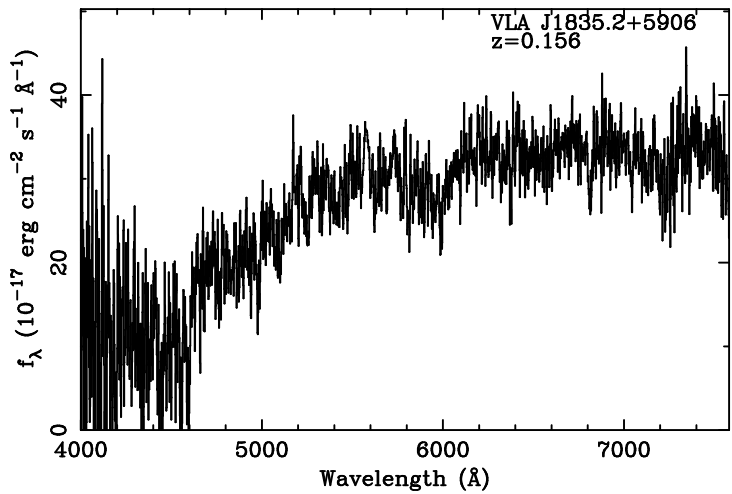
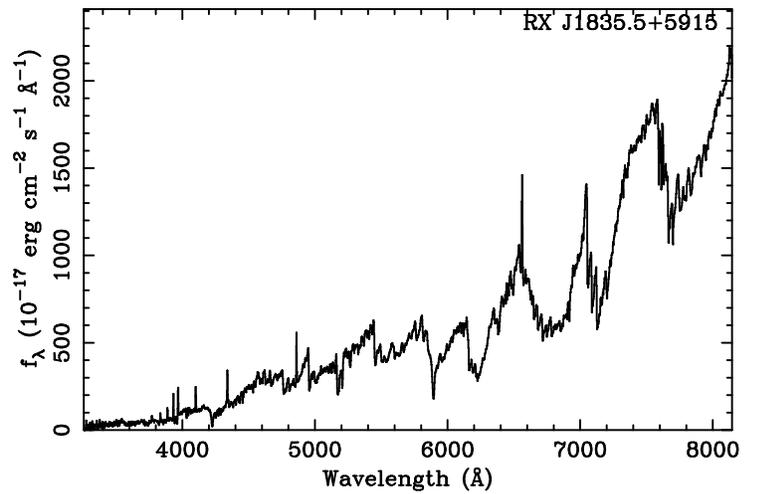
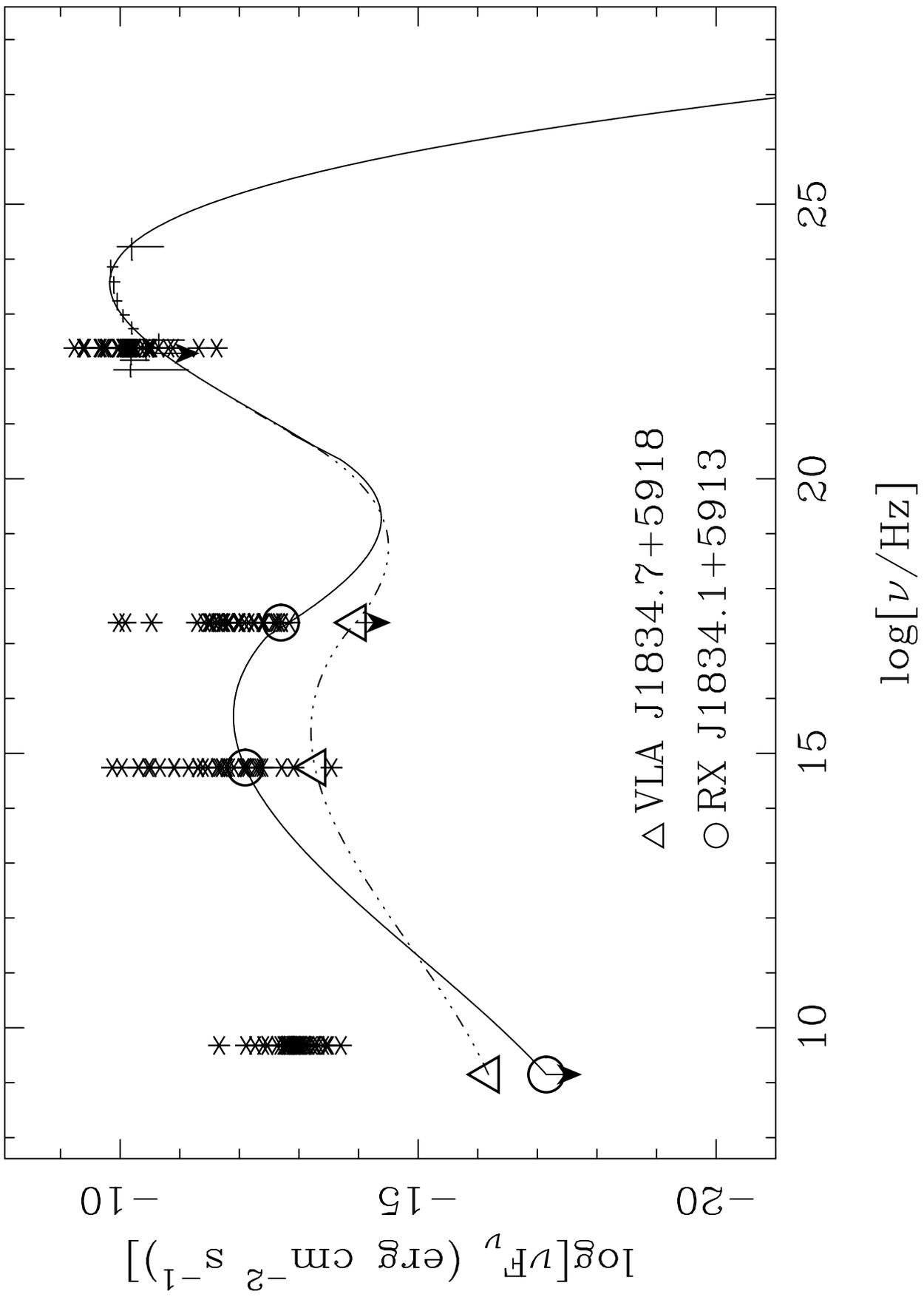
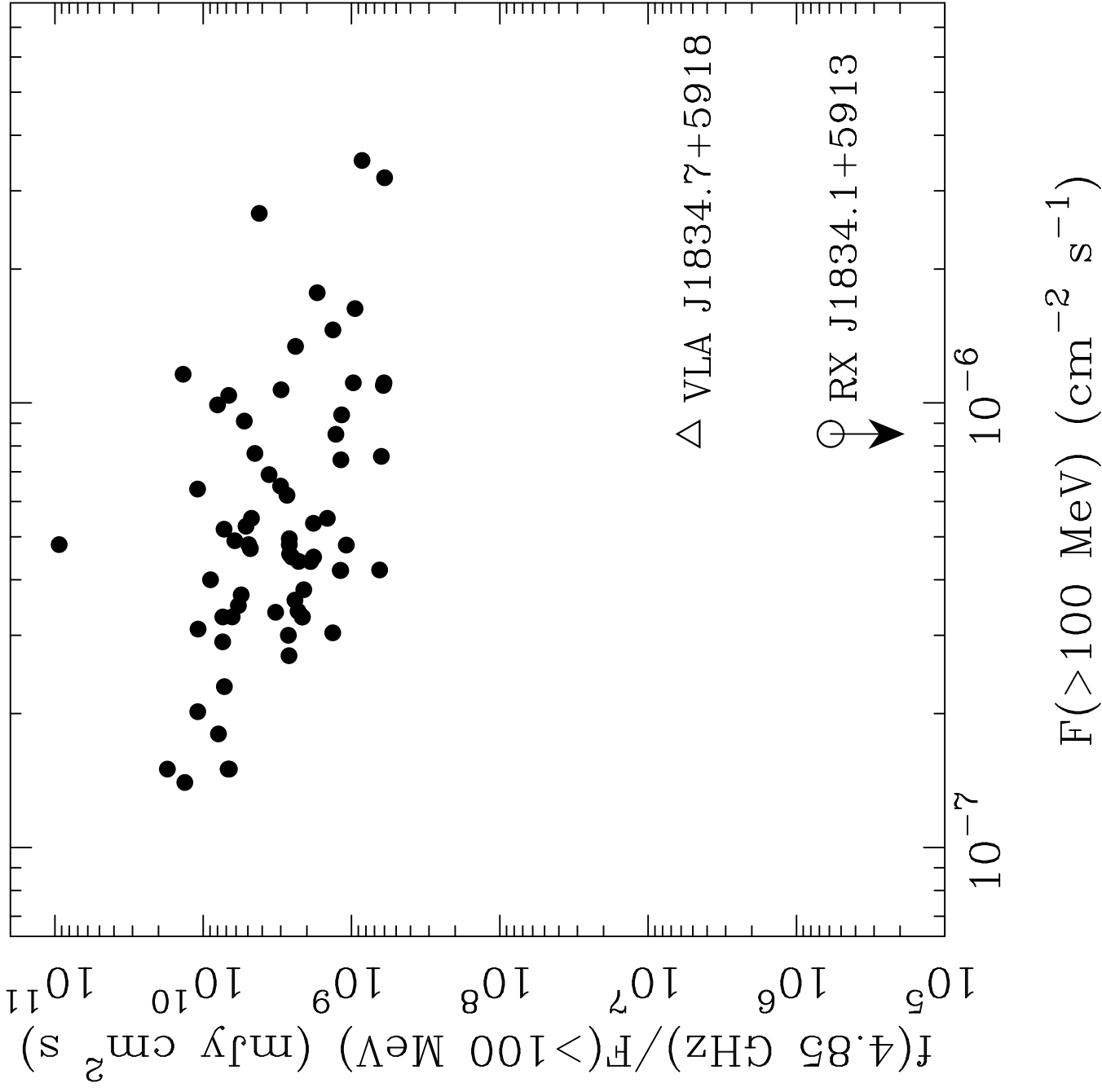
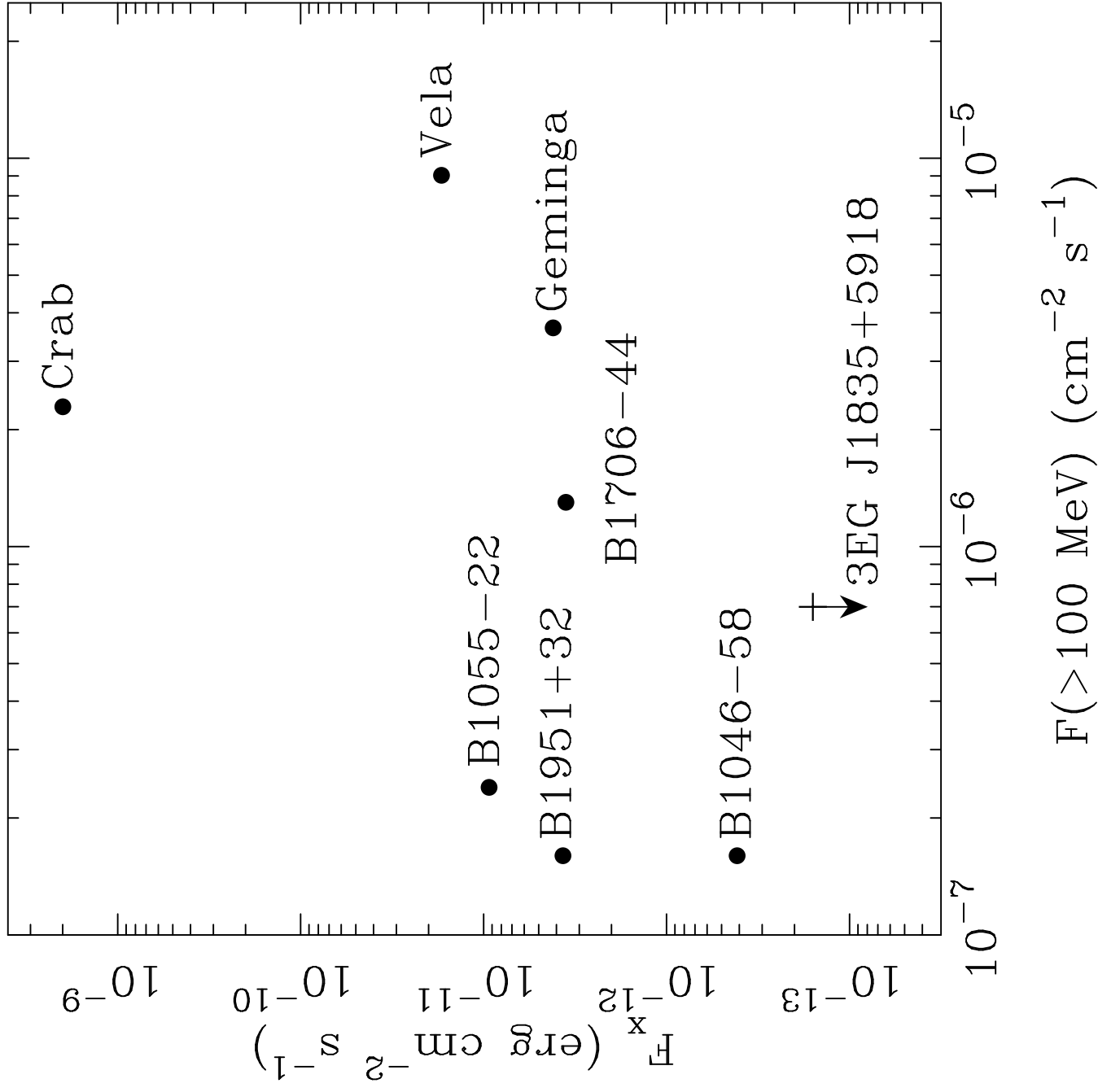
\end{document}